\documentclass[showpacs,prl,twocolumn,aps]{revtex4}
\usepackage{graphicx}
\usepackage{psfrag}
\usepackage{bm}
\begin{document}
\title{Hohenberg Theorem in Position Space}
\title{Existence of Long-Range Order in a  Trapped Bose Gas} 
\title{Existence of Long-Range Order for Trapped Interacting Bosons}
\author{Uwe R. Fischer}
\affiliation{Department of Physics, 
University of Illinois at Urbana-Champaign\\
1110 West Green Street, Urbana, Illinois 61801-3080}



\begin{abstract}                
We derive an inequality governing ``long range'' order for 
a localized Bose-condensed state, relating 
the condensate fraction at a given temperature 
with effective curvature radius of the condensate and total 
particle number. For the specific example of a one-dimensional,
harmonically trapped dilute Bose condensate, 
it is shown that the inequality gives an explicit upper 
bound for the Thomas-Fermi condensate size which may be tested in current 
experiments. 
\end{abstract}

\maketitle
\newcommand{\MF}{{\large{\manual META}\-{\manual FONT}}}
\newcommand{\manual}{rm}        
\newcommand\bs{\char '134 }     


The classic Hohenberg theorem \cite{Hohenberg}
employs the fact that in the thermodynamic limit of an infinite system 
of {\em interacting} particles of {bare} mass $m$, the relation 
($\hbar = k_{\rm B} =1$) 
\begin{equation}
N_{\bm k} \ge -\frac12 + \frac{m T}{k^2} \frac{n_0}n \label{Theorem}
\end{equation}
for the occupation numbers $N_{\bm k}=< \hat b_{\bm k}^\dagger \hat b_{\bm k} >$ 
of the plane wave states enumerated by ${\bm k}$ ($\ne \bm 0$) 
holds (the Bogoliubov $1/k^2$ theorem \cite{Bogoliubov}). 
The angled brackets here and in what follows indicate a thermal
 ensemble (quasi-)average \cite{Bogoliubov}; $n_0$ and
$n$ are the condensate density and total density, respectively. 
The relation (\ref{Theorem}) entails that in one and two dimensions, 
a macroscopic occupation of a single state, the {condensate}, 
is {impossible}:
The inequality leads to a 
contradiction in dimension $D\le 2$ due to the (infrared) 
divergence of the wave vector space integral of (\ref{Theorem}), which 
determines the number density of non-condensate atoms.  
Physically, long range thermal fluctuations destroy the coherence expressed
by the existence of the condensate.

The question of the applicability of the Hohenberg theorem to 
systems displaying long range order has renewed interest 
for trapped Bose-condensed vapors of reduced dimensionality, which can now be
realized using various techniques \cite{Ketterle1D2D,PhaseFluctuation,Haensel,OttMicro,Lattice1D}. In these systems, it is possible to investigate in a controlled
manner the influence of finite extension in different spatial directions
on the existence of a condensate, i.e. a macroscopically occupied 
single state created by $\hat b^\dagger_0$.  
For the {\em non-interacting} case, 
it is not difficult to show that there
can exist a macroscopic occupation of one or several states
in a Bose-condensed, trapped vapor of any dimension, as long 
as particle numbers remain finite, simply by evaluating the Bose-Einstein 
sums for the occupation numbers (see, e.g., \cite{KetterlevanDruten}).
The difficulty comes in when interaction is turned on. An interacting
gas has a behavior increasingly different from the ideal gas the 
lower the dimension of the system \cite{LiebLiniger,1D2D3DGases}. 
Furthermore, the classification of excited non-condensate
states by plane waves is not the suitable one in a trapped gas: 
The condensation, in the limit of large total particle number $N$,  
takes place  primarily 
into a single particle state in co-ordinate space \cite{MullinJLTP97},
and condensate and total densities have (in principle arbitrary) 
spatial dependence, $n_0\rightarrow n_0 ({\bm r})$, $n \rightarrow n({\bm r})$.
While a recent proof by Lieb and Seiringer shows that {ground state} 
condensation is, in the thermodynamic limit,
100\% into the state that minimizes the Gross-Pitaevski\v\i\/
energy functional in three as well as in two dimensions \cite{LiebSeiringer}, 
the question of the existence of a condensate 
for general energy functionals,  
at finite temperatures and, 
in particular,  in the generic one-dimensional (1D) case remains open.

In what follows, we will take account of 
the phenomenon of spatially localized Bose-Einstein condensation,  
by deriving an inequality analogous to %
the integral of (\ref{Theorem}), 
with no explicit dependence on any Hamiltonian 
which has velocity independent interaction and trapping potentials. 
At a given finite temperature, 
the maximally allowed condensate fraction is related to an 
effective curvature radius of the condensate and the total 
particle number.
The inequality thus allows for concrete statements on the limiting size
of quantum coherent systems of reduced dimensionality, and their
realizability for given condensate and total density 
distributions.
As an application to a specific example, we consider  
a 1D harmonically trapped, dilute Bose-Einstein condensate, 
and it is shown that the relation leads to bounds on the parameters
of a Thomas-Fermi cloud which may be verified in present experiments.

Our analysis is based upon the following decomposition of field 
and density operators into condensate and non-condensate parts: 
\begin{eqnarray}
\hat \Phi ({\bm r})& = & \Phi_0({\bm r}) \hat b_0 + \delta \hat \Phi({\bm r})
\nonumber\\
\hat \rho ({\bm r}) & = & 
|\Phi_0({\bm r})|^2\, \hat b_0^\dagger \hat b_0 
+ \delta \hat\rho ({\bm r})\nonumber \\
\delta \hat \rho ({\bm r}) & = &\Phi_0^*({\bm r}) \,
\hat b_0^\dagger \, \delta \hat \Phi ({\bm r}) 
+ {\rm h.c.}
+ \delta \hat \Phi^\dagger({\bm r})\,\delta \hat \Phi ({\bm r}), 
\label{deltarho}
\end{eqnarray} 
where $\Phi_0({\bm r})$ 
is the 
single particle 
condensate wave function (normalized to unity) 
and $N_0 |\Phi_0 ({\bm r})|^2$ is the condensate density. 
The commutation relation 
\begin{eqnarray}
\left[ \delta \hat\Phi ({\bm r}), \delta 
\hat \Phi^\dagger ({\bm r}') \right] 
& =  &
\delta ( {\bm r}- {\bm r}') - \Phi_0({\bm r}) \Phi^*_0({\bm r}'),\label{comm}
\end{eqnarray}
is obtained from the canonical commutation relations for the quantum 
fields $\hat\Phi ({\bm r}), \hat\Phi^\dagger ({\bm r})$.  
As a consequence of this commutator, 
the mixed response (the ``anomalous'' commutator) 
is taking the nonlocal form
\begin{eqnarray}
\left< \left[\hat \rho ({\bm r}), \delta\hat\Phi
({\bm r}')\right]\right>  
\nonumber\\& &  \hspace{-7em}= 
{\sqrt{N_0}}\,\Phi_0({\bm r})
\left[ |\Phi^*_0({\bm r}) \Phi_0 ({\bm r}')-
\delta ( {\bm r}- {\bm r}')\right] 
. \label{PhiOCommutator}
\end{eqnarray}
Here, {\em after} carrying out the commutator, 
$<\hat b_0 >=\sqrt{N_0}= <\hat b^\dagger_0>$ 
has been used, where this assignment is valid to 
$O(1/\sqrt{N_0})$.
The first term in the brackets on the RHS is due to the second term in the 
commutator (\ref{comm}). 
It {\em vanishes} if one takes the Bogoliubov prescription that 
$\hat b_0$ and $\hat b^\dagger_0$ be replaced by c-numbers. It is 
a well-established fact that this (standard) Bogoliubov approach violates 
particle number conservation \cite{TonyBEC}, 
and it becomes apparent below that neglecting 
the second term on the RHS of (\ref{comm}) 
would lead to numerically strongly different predictions for 
the allowed condensate fraction.

We now make use of this form of the 
mixed response in the Bogoliubov inequality \cite{Bogoliubov}, 
which reads 
\begin{equation}
\frac12\left< \left\{ \hat A, \hat A^\dagger \right\} \right>\ge 
\frac{T \left|\left<[\hat C, \hat A] \right> \right|^2}
{\left<\left[\left[\hat C,\hat H\right],\hat C^\dagger\right] \right>} 
\label{Bogoliubov}
\end{equation}
for any two operators $\hat A$ and $\hat C$, where $T$ is the temperature. 
It is valid for any many-body quantum system, for which 
the indicated thermal (quasi-)averages are well-defined. 
We choose the operators in relation (\ref{Bogoliubov}) 
to be the smeared excitation and total density operators
\begin{eqnarray}
\hat A 
& = & \int d^Dr\, f({\bm r}) \delta \hat \Phi ({\bm r}), 
\label{f(r)def}\\
\hat C 
& = &  \int d^Dr\, g({\bm r}) \hat \rho({\bm r}),
\label{g(r)def}
\end{eqnarray}
where $f({\bm r})$ and $g({\bm r})$ 
are complex regularization kernels.

Next we derive a sum rule for the 
denominator on the RHS
 of (\ref{Bogoliubov}), analogous to the
$f$-sum rule $\int_{-\infty}^{+\infty} 
{d\omega}\,\omega S({\bm k},\omega)
= {N k^2 }/{m}$ 
for the dynamic structure factor \cite{PinesNozieres},  
but in co-ordinate space. Using that 
\begin{eqnarray}
\left<\left[\left[\hat \rho ({\bm r}),\hat H ({\bm r})
\right],\hat \rho ({\bm r}')\right] \right>  
& = & -i \left<\left[\nabla_{{\bm r}}\cdot \mbox{$\hat{\j}$}({\bm r})
,\hat \rho ({\bm r}')\right] \right> 
\nonumber \\
&  &   \hspace{-11em} = 
\frac{1}{2m}
\left< 
\delta ({\bm r}'-{\bm r})  
\left( {\hat \Phi}^\dagger ({\bm r}') \Delta_{{\bm r}}
{\hat \Phi} ({\bm r}) + {\rm h.c.}\right)
\right. 
\nonumber  \\
& & \hspace{-9em}  \qquad 
\left. -\Delta_{{\bm r}} \delta ({\bm r}'-{\bm r}) 
\left( {\hat \Phi}^\dagger ({\bm r}'){\hat \Phi} ({\bm r}) + {\rm h.c.} \right)
\right>,
\label{co-ordSumRule}
\end{eqnarray}
we obtain that the double commutator equals 
\begin{equation}
\left<\left[\left[\hat C,\hat H\right],\hat C^\dagger\right] \right> =  
\frac{1}{m} \int\!\! d^Dr \left(-\Delta_{{\bm r}}
g({\bm r})\right) g^*({\bm r})  n ({\bm r})\,, 
\end{equation}
where 
$
 n ({\bm r}) = < \hat\rho ({\bm r}) > =
< \hat \Phi^\dagger({\bm r})\,\hat \Phi ({\bm r}) >.
$
Here, we used the continuity equation for the 
current density operator  
$\hat{\j}=[{\hat \Phi}^\dagger\nabla{\hat \Phi}-(\nabla{\hat \Phi}^\dagger)
{\hat \Phi}]/2im$, together with the 
Heisenberg equation of 
motion for $\hat \rho $:  
$[\hat \rho ,\hat H ]= 
i \partial_t \hat \rho =-i\nabla_{{\bm r}}\cdot \mbox{$\hat{\j}$}$, 
this last relation being valid for a  
Hamiltonian with no explicit velocity dependence 
in the interaction and external potentials \cite{PinesNozieres}.    

We normalize the kernel $f({\bm r})$ to unity, i.e., 
$\int\!\! d^Dr\, |f({\bm r})|^2 =1$. 
The left hand side of the Bogoliubov inequality (\ref{Bogoliubov})
is then bounded from above, using the Cauchy-Schwarz inequality, as follows
\begin{eqnarray}\hspace*{-1em}
& &\frac12\int\!\! d^Dr\!\! \int \! \! d^Dr' 
{f({\bm r})} {f^*({\bm r}')}
\left< \left\{\delta\hat \Phi ({\bm r}) , \delta\hat \Phi^\dagger ({\bm r}')
 \right\} \right>  \nonumber\\
& & \hspace{1em} \le  
\int\!\! d^Dr'
\left< \delta\hat \Phi^\dagger ({\bm r}') \delta\hat \Phi ({\bm r}')
\right> + \frac a2 
\nonumber\\
& & \hspace{1em} =  
N-N_0 +\frac a2\,,
\end{eqnarray}
where the quantity $a\le 1$ is given by 
$a=1- {\int\!\! d^Dr \int\!\!  d^Dr' 
f({\bm r}) f^*({\bm r}')
 \Phi_0 ({\bm r})  \Phi^*_0 ({\bm r}')}\,,$ 
and where $N - N_0  
= \int\!\! d^Dr < \delta\hat\rho ({\bm r}) > 
=\int\!\! d^Dr 
< \delta \hat \Phi^\dagger({\bm r})\,\delta \hat \Phi ({\bm r}) >$
is the excited number of particles.

Using the anomalous commutator (\ref{PhiOCommutator}), 
the Bogoliubov inequality (\ref{Bogoliubov})
thus may be written in the following form: 
\vspace*{-2em}
\begin{widetext}
\begin{eqnarray}
{N-N_0}  
\ge -\frac a{2}
+{mTN_0}\,
\frac{\left|
\int d^Dr\, g({\bm r}) f({\bm r}) 
\Phi_0 ({\bm r})
- \int d^Dr \int d^Dr'\, 
g({\bm r}) f({\bm r}')
|\Phi_0({\bm r})|^2 \Phi_0 ({\bm r}')^{} 
\right|^2}
{
\int\!\! d^Dr \left(-\Delta_{{\bm r}}
g({\bm r})\right) g^*({\bm r})  n ({\bm r})}
\,. 
\label{FinalInequality}
\end{eqnarray}
\end{widetext}
It is important to recognize that the above relation is explicitly 
independent of the form of the excitation spectrum of the system, due to 
the relation (\ref{co-ordSumRule}). 
In particular, the strength of the 
interaction enters only implicitly in the form of the condensate wave 
function and total density distribution.  
This is in contrast to previous considerations on ``long range'' order
in Bose-Einstein condensates, employing correlation functions 
\cite{1D2D3DGases,HoMa}, where use was made of the excitation 
spectrum of the system, with Bogoliubov-type or WKB approximations.

Now choose the kernels to have the particular form 
\renewcommand{\arraystretch}{1.5}\begin{eqnarray}
f_{\bm k}({\bm r})
& = &\left\{\begin{array}{lr} {\Omega}_0^{-1/2} 
\exp[i{\bm k}\cdot {\bm r}] 
 & \quad ({\bm r} \in {\cal D}_0)  \\
 0 & \quad ({\bm r} \notin {\cal D}_0)
\end{array}\right. \nonumber 
\\
 g({\bm r}) & = & \left\{\begin{array}{lr}
\Phi_0^*({\bm r})& \quad \qquad \qquad\;\,\, ({\bm r} \in {\cal D}_0)
\\
 0 & \quad \qquad \qquad \;\,\,({\bm r} \notin {\cal D}_0)
\end{array}\right.
\label{fgAnsatz}\renewcommand{\arraystretch}{1.0}
\end{eqnarray}
where ${\Omega}_0$ is the volume of the 
domain ${\cal D}_0$ in which $\Phi_0$ has finite support. Hence 
$|g|^2$ is also, automatically, normalized to unity, 
$\int_{{\cal D}_0} d^Dr|g|^2 = 
\int_{{\cal D}_0} d^Dr|\Phi_0|^2=1$.
The choice of $g$ 
is motivated by requiring that it contains only information 
pertaining to (the spatial extension of) the condensate. The kernels 
$f_{\bm k}$ are a {set} of functions in which we vary
$\bm k$ such that the RHS of (\ref{FinalInequality}) is maximized   
and the inequality is assuming its strongest form. 

After introducing the kernels (\ref{fgAnsatz}) into 
the Bogoliubov inequality (\ref{FinalInequality}), we have
our primary result:  
\vspace*{-2em}
\begin{widetext}
\begin{eqnarray}
{N-N_0} 
& \ge & 
-\,\frac{1 - \frac{1}{\Omega_0}\left| \tilde \Phi_0 ({\bm k}) \right|^2}{2}
+\frac{T N_0}{2{\Omega}_0}\,\frac{\left|\tilde n_0 ({\bm k}) 
- \tilde \Phi_0 ({\bm k})\int_{{\cal D}_0} d^Dr \,
\Phi_0^*({\bm r}) |\Phi_0({\bm r})|^2\right|^2}
{\int_{{\cal D}_0} d^Dr\, \Phi_0({\bm r})\left[
-\frac{1}{2m}\Delta_{{\bm r}}\Phi_0^*({\bm r})\right] n ({\bm r})}
. \label{Final}
\end{eqnarray}
\end{widetext}
The Fourier transforms of single particle 
condensate density and wave function are defined
to be 
$\tilde n_0({\bm k}) =\int d^Dr \,
|\Phi_0 ({\bm r})|^2 \exp[i{\bm k}\cdot {\bm r}]$ and 
$\tilde \Phi_0 ({\bm k}) =\int d^Dr\, \Phi_0 ({\bm r}) \exp[i{\bm k}\cdot {\bm r}]$.

We define the {\em effective radius of curvature} of the condensate to be 
\begin{equation}
{R_c}\equiv \sqrt{\frac{N}{\Omega_0}}\! \left(
{\int_{{\cal D}_0}\!\! d^Dr\,  \Phi_0({\bm r})\left[
-\Delta_{{\bm r}}\Phi_0^*({\bm r})\right] n ({\bm r})}\!
\right)^{-1/2}\!.
\end{equation}
According to the above formula, 
$R_c$ is obtained by weighing a quantity proportional to 
the kinetic energy density of the condensate 
with $n({\bm r})/(N/\Omega_0)$, i.e., the 
local density relative to the average density, 
and finally integrating
over the domain ${\cal D}_0$. 
Using this definition, 
relation (\ref{Final}) reads 
\begin{equation}
\frac{1-F}F \ge \frac1N\frac{2\pi R_c^2}{\lambda_{\rm dB}^2} \,{\cal C} ({\bm k}) 
-\frac{1}{2N_0}\left(1 - 
|\tilde \Phi_0 ({\bm k})|^2/\Omega_0\right), 
\label{RcCF}
\end{equation}
where the functional ${\cal C} ({\bm k})$ is given by 
\begin{equation}
{\cal C} ({\bm k})= \left|\tilde n_0({\bm k}) -\tilde \Phi_0 ({\bm k})
\int_{{\cal D}_0} d^Dr\, \Phi_0^*({\bm r}) |\Phi_0({\bm r})|^2 
\right|^2, \label{Ckdef}
\end{equation} 
and the condensate fraction $F= N_0/N$; 
the de Broglie thermal wavelength $\lambda_{\rm dB} = \sqrt{2\pi /mT}$.
We stress that the value of ${\cal C} ({\bm k})$ is strongly 
reduced as a consequence of the commutation relation (\ref{comm}), which 
causes the second term under  
the  square in (\ref{Ckdef}). 

The relation (\ref{RcCF}) implies that for temperatures close to zero, 
such that $F R_c^2/\lambda_{\rm dB}^2$ remains large 
(in which case the second term in  (\ref{RcCF}) is negligible),  
the approach of $N{\cal C}^{-1} (1-F) 
= {\cal C}^{-1} (N-N_0)\propto T^\alpha$ 
to zero with a power law has to fulfill $\alpha\le 1$ for complete 
Bose-Einstein condensation into a localized 
single state ${\Phi_0}({\bm r})$ to be possible.
This statement holds for 
arbitrary strength and form of the interaction, in any spatial dimension.

The above relations (\ref{Final}) and (\ref{RcCF}) are general. 
The strongest result, i.e., constraint on the system parameters 
we may expect for a 1D system, in analogy to the original 
Hohenberg theorem (the $1/k$ infrared divergence of the 
integral of (\ref{Theorem}) in one dimension).  
To demonstrate the meaning of the relation (\ref{Final})
explicitly, we thus now proceed by considering the example of
axially symmetric,   
harmonically trapped gases in one dimension, in the currently
experimentally accessible Thomas-Fermi limit.

\begin{figure}[htbp]
\psfrag{x}{\huge $x=kZ_{\rm TF}$}
\psfrag{100*C(x)}{\huge $100\times {\cal C}(x)$}
\vbox{
\hfil
\scalebox{0.48}{\includegraphics{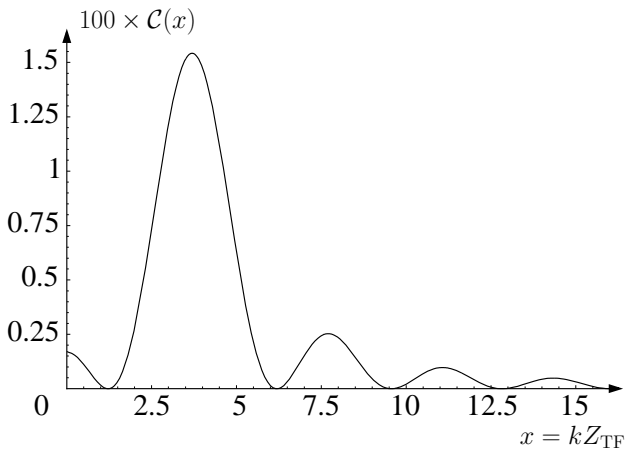}}
\hfil
}
\caption{
The function ${\cal C}({\bm k})$ on the RHS of (\ref{RcCF}) in the 
one-dimensional Thomas-Fermi case,  from Eq.\,(\ref{C(x)}). 
}
\label{C:sketch}\vspace{-1em}
\end{figure}

Consider the Thomas-Fermi wave function
\begin{equation}
\Phi_0 (z) = \sqrt{\frac{n^0_{\rm TF}}{N}} 
\left(1-\frac{z^2}{Z_{\rm TF}^2} \right)^{1/2}.
\label{1DZTF}
\end{equation}
This mean-field form of $\Phi_0$ is  
valid if the 1D scattering length fulfills the {\em strong
coupling} condition $n|a_{\rm 1D}| \gg 1$, and 
the Thomas-Fermi parameters are 
$Z_{\rm TF}= (3N d_z^4/|a_{\rm 1D}|)^{1/3}$, 
$n_{\rm TF}^0 = [(9/64) N^2|a_{\rm 1D}|/d_z^4]^{1/3}$ \cite{Dunjko};  
the quantities 
$d_{\perp,z} = (m\omega_{\perp,z})^{-1/2}$ are the harmonic oscillator lengths.
In the quasi-3D scattering
limit, which has transverse length scale  
$d_\perp\gg a_s$, 
 the 1D scattering length is given by 
$a_{\rm 1D} = -(d_\perp^2/2a_s) [1- C a_s/d_\perp]$, 
with $C=1.4603$ \cite{Dunjko}. 
We neglect the difference between $n ({\bm r})$ and 
$ N |\Phi_0 ({\bm r})|^2$, i.e., 
take the limit of both sides of (\ref{Final}) to linear order in $1-F$. 
Evaluating the elementary integrals involved, 
we obtain that ($x\equiv k Z_{\rm TF}$) 
\begin{equation}
{\cal C} (x)
= \left| \frac{3}{x^2} \left(\frac{\sin x}{x} -\cos x \right)
-\frac{27\pi^2}{128}\frac{J_1(x)}x  
\right|^{2}, \label{C(x)}
\end{equation}
where $J_1(x)$ is a Bessel function of the first kind.
We see from Fig.\,\ref{C:sketch} that 
the function ${\cal C}$ is strongly peaked at its global maximum 
$k_m \simeq 3.7 Z_{TF}^{-1}$ ($\lambda_m \simeq 1.7 Z_{\rm TF})$,  
where ${\cal C}(k_m)\simeq 1.54\times 10^{-2}$. 
Using that $R_c = 4Z_{\rm TF}/3\sqrt{2\pi}$, 
and neglecting the second term on the RHS of (\ref{RcCF}), 
we obtain at $k=k_m$  
\begin{eqnarray}
\frac{Z_{\rm TF}}{\lambda_{\rm dB}} \le 6.0
\, \sqrt{{N-N_0}} 
\,. \label{1DZN}
\end{eqnarray}

We compare the above relation 
with the experiment on a 
1D $^{23}\!$Na condensate in \cite{Ketterle1D2D}, where
the (relatively moderate) parameters were $a_s = 2.8 \times 10^{-3}\,
\mu $m, $d_z=11.2\,\mu$m ($\omega_z/2\pi =3.5 $\,Hz), 
$d_\perp=1.15\,\mu$m, and $N \sim 1.5\times 10^4$, which result in 
$Z_{\rm TF} \simeq 150\,\mu$m.
Inequality (\ref{1DZN}) becomes $Z_{\rm TF}/\lambda_{\rm dB} 
\lesssim 7.3\times 10^2\sqrt{1-F}$. 
Temperatures in this experiment have been of the order $T\sim O(100\, $nK)
\cite{KetterlePrivComm}, which gives $\lambda_{\rm dB}\simeq 1 \,\mu$m
and $Z_{\rm TF}/\lambda_{\rm dB}\simeq 1.5 \times 10^2$.
The parameters of \cite{Ketterle1D2D} are thus consistent with  (\ref{1DZN}), 
provided $F$ is not too close to unity \cite{idealgas}.  
Note that (\ref{1DZN}) would be {\em inconsistent} 
with the parameters in that experiment, 
were it not for the commutation relation 
(\ref{comm}), due to imposing the canonical commutation relations for 
the {\em total} quantum field. Neglecting the second term on the RHS
of (\ref{comm}) leads to a decrease of the RHS of (\ref{1DZN}) by about 
one order of magnitude. In this sense, particle number conservation, which is
violated by the standard Bogoliubov prescription, is necessary for the 
condensate to exist \cite{Olinto}. 

A further decrease of the aspect ratio $\omega_z/\omega_\perp = 
d_\perp^2/d_z^2$ 
down to values of order $10^{-3}$ 
and lower is achievable, e.g.,  
in optical lattices \cite{Lattice1D}, 
so that the condition (\ref{1DZN}) on the trap parameters
should be experimentally verifiable within present technology.
When (\ref{1DZN}) ceases to be fulfilled, the system has to 
enter into a new (non-Thomas-Fermi) state.
For very low densities, 
a ``Tonks gas'' (1D gas of impenetrable particles) 
may be formed, which has no condensate 
\cite{1D2D3DGases,Dunjko,GirardeauWright}. 
Another possibility 
is a condensate consisting of (overlapping) phase coherent droplets 
\cite{1Dsupersolid}, resulting in a reduced value of $R_c$. 

The primary result of the present investigation, inequality (\ref{Final}),
holds under the generic conditions 
that the Bogoliubov inequality (\ref{Bogoliubov}) is valid 
and that the potentials in the Hamiltonian are independent 
of particle velocities.  
It is, furthermore, to be emphasized that the application of (\ref{Final}) 
is by no means limited to (real) ground state forms of $\Phi_0$.
It is also possible to employ this relation 
to examine existence conditions for excited state condensates 
with complex $\Phi_0$, like single vortices or vortex lattices.  
While (\ref{Final}) 
certainly cannot guarantee
the existence of a given $\{N_0, \Phi_0({\bm r}), n({\bm r}) \}$ state, 
it can rule out models for trapped Bose-condensed gases in various 
spatial dimensions which are inconsistent with the 
Bogoliubov inequality (\ref{Bogoliubov}).

I am indebted to Tony Leggett for many helpful
discussions, which provided a major inspiration for the present work. 
The author acknowledges support by the {\it Deutsche 
Forschungsgemeinschaft}  (FI 690/2-1).  
This research was also supported in part by
NSF Grant DMR99-86199.


\begin{thebibliography}{499}
\bibitem{Hohenberg} P.\,C. Hohenberg,
Phys. Rev. {\bf 158}, 383 
(1967).
\bibitem{Bogoliubov} N.\,N. Bogoliubov, {\em Selected Works, Part II: 
Quantum and Statistical Mechanics} (Gordon and Breach, 1991). 
\bibitem{Ketterle1D2D} A. G\"orlitz {\it et al.}, 
Phys. Rev. Lett. {\bf 87}, 130402 (2001).
\bibitem{PhaseFluctuation} S. Dettmer {\it et al.}, 
Phys. Rev. Lett. {\bf 87}, 160406 (2001). 
\bibitem{Haensel} W. H\"ansel, P. Hommelhoff, T.\,W. H\"ansch, and J. Reichel, 
Nature {\bf 413}, 498 
(2001). 
\bibitem{OttMicro} H. Ott, J. Fort\'agh, G. Schlotterbeck, A. Grossmann, and 
C. Zimmermann, 
Phys. Rev. Lett. {\bf 87}, 230401 (2001). 
\bibitem{Lattice1D} M. Greiner, I. Bloch, O. Mandel, T.\,W. H\"ansch, 
and T. Esslinger, Phys. Rev. Lett. {\bf 87}, 160405 (2001). 
\bibitem{KetterlevanDruten} W. Ketterle and N.\,J. van Druten,  
Phys. Rev. A {\bf 54}, 656 
(1996); N.\,J. van Druten and W. Ketterle, 
Phys. Rev. Lett. {\bf 79}, 549 (1997).
\bibitem{LiebLiniger} E.\,H. Lieb and W. Liniger,
Phys. Rev. {\bf 130}, 1605 
(1963).  
\bibitem{1D2D3DGases} 
D.\,S. Petrov, M. Holzmann, and G.\,V. Shlyapnikov, 
Phys. Rev. Lett. {\bf 84}, 2551 
(2000); D.\,S. Petrov, G.\,V. Shlyapnikov, 
and J.\,T.\,M. Walraven, {\it ibid.} {\bf 85}, 3745 
(2000); 
{\it ibid.} {\bf 87}, 050404 (2001). 
\bibitem{MullinJLTP97} W.\,J. Mullin, 
J. Low Temp. Phys. {\bf 106}, 615 
(1997).
\bibitem{LiebSeiringer} E.\,H. Lieb and R. Seiringer, 
Phys. Rev. Lett. {\bf 88}, 170409 (2002).
\bibitem{TonyBEC} A.\,J. Leggett, Rev. Mod. Phys. {\bf 73}, 307 
(2001).
\bibitem{PinesNozieres} D. Pines and  
P. Nozi\`eres, {\it The Theory of Quantum 
Liquids: Volume I} (W.\,A. Benjamin, 1966). 
\bibitem{HoMa} T.-L. Ho and M. Ma,
J. Low Temp. Phys. {\bf 115}, 61 
(1999).
\bibitem{Dunjko} V. Dunjko, V. Lorent, and M. Olshani\v\i\/,
Phys. Rev. Lett. {\bf 86}, 5413 (2001); 
M. Olshani\v\i\/, {\it ibid.} {\bf 81}, 938 (1998).
\bibitem{KetterlePrivComm} J.\,M. Vogels and W. Ketterle 
(private communication).
\bibitem{idealgas} Observe that if ${N-N_0}$ is close to the ideal gas 
prediction ${N-N_0}=2\pi (d^2_z/\lambda^2_{\rm dB}) \ln[2N]$ 
 \cite{KetterlevanDruten}, relation (\ref{1DZN}) becomes {independent} 
of temperature, and leads to the lower bound for the aspect ratio,\,
${\omega_z}/{\omega_\perp} 
\, 
\ge N {\tilde a_s }/ 
(5.8\times 10^2 d_z \ln^{3/2} [2N]) \,$, 
where ${\tilde a_s} =a_s/|1- C a_s/d_\perp|$. In the thermodynamic limit,
$d_z \rightarrow \infty$, $N\rightarrow \infty$, 
with $ N/d^2_z $ fixed,  condensation is then ruled out.  
This statement, however, should be taken {\it cum grano salis}: 
In the homogeneous case, the crossover from interacting 
to noninteracting has been shown to be nonanalytic 
for a contact interaction potential of finite strength \cite{LiebLiniger}, 
and ${N-N_0}$ may grow stronger than only logarithmically with $N$. 
\bibitem{Olinto} The point that the standard non-number-conserving 
Bogoliubov approach, 
when applied to the Bogoliubov inequality,  
is inconsistent with existing Bose condensates 
has also been made by A.\,C. Olinto, Phys. Rev. A {\bf 64}, 033606 (2001). 
However, it was claimed that, as a consequence,
there is a need to modify the canonical commutation relations, leading
to explicit dependence of the Bogoliubov inequality on particle
interaction.
It was shown here that this modification is {\em not} necessary, 
and that explicit dependence on interaction does not need to occur.
\bibitem{GirardeauWright} M.\,D. Girardeau and E.\,M. Wright,
Phys. Rev. Lett. {\bf 87}, 210401 (2001).  
\bibitem{1Dsupersolid} S. Giovanazzi, $\!$ D. O'Dell, $\!$ and G. Kurizki, 
Phys. Rev. Lett. {\bf 88}, 130402 (2002) discussed a ``supersolid'' 
phase for a self-bound condensate,  
in the presence of an additional laser-induced interaction potential.
\end{thebibliography}
\end{document}